\documentclass{jfm}

\ifCUPmtlplainloaded \else
  \checkfont{eurm10}
  \iffontfound
    \IfFileExists{upmath.sty}
      {\typeout{^^JFound AMS Euler Roman fonts on the system,
                   using the 'upmath' package.^^J}%
       \usepackage{upmath}}
      {\typeout{^^JFound AMS Euler Roman fonts on the system, but you
                   dont seem to have the}%
       \typeout{'upmath' package installed. JFM.cls can take advantage
                 of these fonts,^^Jif you use 'upmath' package.^^J}%
      }
  \else
  \fi
\fi

\ifCUPmtlplainloaded \else
  \checkfont{msam10}
  \iffontfound
    \IfFileExists{amssymb.sty}
      {\typeout{^^JFound AMS Symbol fonts on the system, using the
                'amssymb' package.^^J}%
       \usepackage{amssymb}%

      }{}
  \fi
\fi

\ifCUPmtlplainloaded \else
  \IfFileExists{amsbsy.sty}
    {\typeout{^^JFound the 'amsbsy' package on the system, using it.^^J}%
     \usepackage{amsbsy}}
    {\providecommand\boldsymbol[1]{\mbox{\boldmath $##1$}}}
\fi

\newsavebox{\astrutbox}
\sbox{\astrutbox}{\rule[-5pt]{0pt}{20pt}}

\newcommand\etal{\mbox{\textit{et al.}}}

\newcommand\eg{e.g.\ }

\title[LES of passive scalar: a systematic approach]{ 
Large-Eddy Simulation closures of passive scalar turbulence: a systematic approach}

\author[M.~Martins Afonso, A.~Celani, R.~Festa and A.~Mazzino ]%
{M. \ns M\ls A\ls R\ls T\ls I\ls N\ls S \ns A\ls F\ls O\ls N\ls S\ls O$^1$,\ns  A. \ns C\ls E\ls L\ls A\ls N\ls I$^2$, \break
\ns R. \ns F\ls E\ls S\ls T\ls A$^1$ \and A. \ns M\ls A\ls Z\ls Z\ls I\ls N\ls O$^{3,1}$}

\affiliation{$^1$ INFM--Department of Physics, University of Genova, I--16146
Genova, Italy\\[\affilskip]
$^2$ CNRS, INLN, 1361 Route des Lucioles, 06560 Valbonne, France\\[\affilskip]
$^3$ ISAC/CNR, Lecce Section, I--73100 Lecce, Italy}

\pubyear{2003}
\volume{???}
\pagerange{1--9}
\date{?? and in revised form ??}
\usepackage{graphicx}

\begin{document}

\maketitle

\begin{abstract}
The issue of the parameterization of small scale (``subgrid'')
turbulence is addressed in the context of passive scalar
transport. We focus on the Kraichnan advection model
which lends itself to the analytical investigation
of the closure problem. We derive systematically the
dynamical equations which rule the evolution of the 
coarse-grained scalar field. At the lowest-order approximation
in $l/r$, $l$ being the characteristic 
scale of the filter defining the coarse-grained scalar field and
$r$ the inertial range separation,
we recover the classical eddy-diffusivity parameterization of small scales.
At the next-leading order a dynamical closure  is obtained.
The latter outperforms the classical model and is therefore a 
natural candidate for subgrid modelling of scalar transport in 
generic turbulent flows.
\end{abstract}

%
%
\section{Introduction}

One of the most striking characteristics of hydrodynamic
turbulence is the presence of a wide range of active length and
time scales. These scales are strongly and nonlinearly coupled, a fact that
makes analytical approaches, at best, impractical. The
situation does not look better for direct numerical simulations
of turbulent systems: to fully resolve a turbulent flow requires
approximately $(L/\eta)^{3/4}$ grid points in each spatial direction (see, for
example, \cite[Frisch 1995]{F95}), $L$ and $\eta$ being
the integral scale and the dissipation scale respectively. In the
atmosphere, for instance, the ratio $L/\eta$ may become of
the order of $10^{10}$ ($\eta\sim 10^{-3}\;m$ and $L\sim 10^{7}\;m$)
thus requiring the dynamical description of $10^{22}$ degrees of
freedom. This remains, up to now and probably also in the near
feature, a prohibitive task.

To overcome the problem, ``coarse-grained'' versions of the original
hydrodynamic equations are often considered in order to describe
large-scale features of the original full system. The Large-Eddy
Simulation (LES) technique is probably the most popular example 
(\cite[Meneveau \& Katz 2000]{MK00}).
The success of such strategy is however strongly dependent on the
realism of the description of small scales in terms of
the large, explicitly resolved, scales. 
The problem of representing small unresolved scales in the absence of scale
separation --  the long-known
{\it closure problem} -- 
attracts a great deal of attention in many
domains ranging from geophysics to engineering (\cite[McComb 1992]{MC00}), and
is one among the many challenges of
turbulence theory.

Our goal here is to shed some light on this 
aspect within the context of scalar turbulence where  considerable
progresses have been achieved in the last few years 
(\cite[Shraiman \& Siggia 2000]{SS00}, \cite[Falkovich \etal\ 2001]{FGV01}).
For this purpose we will consider a particular model of scalar transport 
(\cite[Kraichnan 1968]{K68}, \cite[Kraichnan 1994]{K94}) 
where the LES strategy can be 
formulated and the problem of relating unresolved scales 
to resolved ones can be successfully attacked analytically.

In this respect,
the Kraichnan model has some characteristics of paramount importance:
\begin{itemize}
\item[\mbox{--}] Exact expressions for relevant 
statistical observables can be derived from first principles, that is 
from Eq.~(\ref{fp}): this amounts to saying that the observables for the 
``fully resolved case'' are known. An example is the expression
 (\ref{s2}) for the second order scalar structure function,
an observable tightly related to the Fourier spectrum of the scalar field;
\item[\mbox{--}] Closures for the  large-scale dynamics can be derived 
in a systematical way (see~\S\,\ref{systematic}),
and their predictions can be analytically checked against the exact solution.
\end{itemize}
Those features make the Kraichnan model an ideal playground
for studying LES closures. 

\section{LES closures for passive scalar turbulence}
Scalar transport is governed by the
advection-diffusion equation
\begin{equation}
\label{fp}
\partial_t \theta +\boldsymbol{v}\cdot\nabla\,
\theta =\kappa_0\Delta \theta +f\;,
\end{equation}
describing the
evolution of a  passive scalar field $\theta(\boldsymbol{x},t)$ --
\eg temperature when buoyancy effects are negligible -- advected by
 an incompressible velocity field $\boldsymbol{v}(\boldsymbol{x},t)$.
Scalar fluctuations are injected into the system at the large scale
$L$ by the forcing term $f$. Dissipation takes on at small
scales $\eta$ due to the
molecular diffusivity $\kappa_0$.

The coarse-grained scalar and velocity fields, denoted by $\tilde{\theta}$
and $\tilde{\boldsymbol{v}}$ are obtained by convolving the original, fully
resolved, fields with a filter $G_l$ with characteristic scale $l$  
($\eta \ll l \ll L$)
\begin{equation}
\tilde{\theta}(\boldsymbol{x},t)=\int G_l(\boldsymbol{x}-\boldsymbol{x}')
\theta(\boldsymbol{x}',t)\,d\boldsymbol{x}'\;,
\end{equation}
\begin{equation}
\tilde{\boldsymbol{v}}(\boldsymbol{x},t)=\int G_l(\boldsymbol{x}-\boldsymbol{x}')
\boldsymbol{v}(\boldsymbol{x}',t)\,d\boldsymbol{x}'\;.
\end{equation}
The equation for $\tilde{\theta}$ derived from  Eq.~(\ref{fp}) is:
\begin{equation}
\partial_t \tilde{\theta}+\tilde{\boldsymbol{v}}\cdot\nabla\,
\tilde{\theta}=\kappa_0\Delta\tilde{\theta} +
\tilde{f} -\left( {\cal L} + \tilde{\cal S} \right ) 
\label{fp-les}
\end{equation}
where ${\cal L}$ is analogous to the Leonard stress
\begin{equation}
{\cal L}\equiv \widetilde{\tilde{\boldsymbol{v}}\cdot\nabla\,\tilde{\theta}}
-\tilde{\boldsymbol{v}}\cdot\nabla\,\tilde{\theta} \;,
\label{leo-stress}
\end{equation}
and ${\cal S}$ is defined as
\begin{equation}
{\cal S}\equiv \boldsymbol{v}'\cdot\nabla\,\tilde{\theta}+
\boldsymbol{v}'\cdot\nabla\,\theta '+ \tilde{\boldsymbol{v}}\cdot\nabla\,\theta'\;.
\end{equation}
The small-scale fields $\theta'$ and $\boldsymbol{v}'$ 
are defined as $\theta '\equiv \theta-\tilde{\theta}$  
and $\boldsymbol{v}'\equiv \boldsymbol{v}-\tilde{\boldsymbol{v}}$.

The purpose of LES closures is to express ${\cal L}$ and 
$\tilde{\cal S}$ in terms of the large scale fields
$\tilde{\theta}$ and $\tilde{\boldsymbol{v}}$. Once this goal is
accomplished, Eq.~(\ref{fp-les}) can be numerically integrated
on a mesh of spacing $l$, rather than $\eta$ as it
should be required for the integration of the full system (\ref{fp}),
with an enormous gain in memory and CPU time requirements.

Unfortunately, no general closed expression for 
${\cal L}$ and $\tilde{\cal S}$ in terms of $\tilde{\theta}$
and $\tilde{\boldsymbol{v}}$ is available.
A remarkable exception is the case where there is a marked scale 
separation between velocity and scalar length and timescales. It is then
possible to show (see, \eg, \cite[Biferale \etal\ 1995]{BCVV95},
\cite[Mazzino 1997]{M97}) that
the effect of unresolved scales is just the renormalization
of  the molecular diffusion coefficient $\kappa_0$ 
to an enhanced eddy-diffusivity
$\kappa^{\mbox{\tiny{eff}}}$ (generally speaking, an eddy-diffusivity
tensor $\kappa^{\mbox{\tiny{eff}}}_{ij}$).
General expressions of the eddy-diffusivity as a function of the flow 
properties do not exist, and in most cases 
 $\kappa^{\mbox{\tiny{eff}}}$ can be determined only numerically.

Here, our aim is to consider the challenging situation where there is no scale
separation between velocity and scalar and explore, in such context, 
the existence of effective equations for $\tilde{\theta}$.

Our procedure to derive the closed 
coarse-grained dynamical equations is the following:
\begin{itemize}
\item[({\it i\/})] starting form first principles, 
that is from Eq.~(\ref{fp}), we take advantage of the 
peculiarity of the Kraichnan model to derive
an exact, yet unclosed, statistical equation for the correlation function of 
the filtered scalar field 
$\langle \tilde{\theta}(\boldsymbol{x},t)  \tilde{\theta}(\boldsymbol{x}+\boldsymbol{r},t)\rangle $; 
\item[({\it ii\/})] the statistical averages appearing in this equation,
which involve small-scale 
fields, are then expressed in terms of correlations of large-scale fields only,
at a given order of approximation in $l/r$; 
\item[({\it iii\/})] from the statistically closed equations we consistently 
infer the dynamical closed equations for $\tilde{\theta}$.
\end{itemize}
No supplementary assumptions are made in performing this procedure.
Although the method relies heavily on peculiar characteristics 
of the Kraichnan model, we believe that the results are relevant
to generic passive scalar turbulence as well. Our claim is supported
by several numerical and analytical evidences gathered in the last few
years showing that most of the phenomenology of
scalar turbulence is captured by the Kraichnan model. 
For an exhaustive review on this aspect, see, 
\eg \cite[Falkovich \etal\ (2001)]{FGV01}.

To illustrate the power of our approach we anticipate here the main results
of this paper, postponing their derivation to the following sections.

Carrying out the procedure at the lowest significant order
in $l/r$ yields the following effective equation for the coarse-grained field:
\begin{equation}
\partial_t \tilde{\theta}+\tilde{\boldsymbol{v}}\cdot\nabla\,
\tilde{\theta}=\kappa^{\mbox{\tiny{eff}}}\Delta\tilde{\theta} +
\tilde{f}\qquad \kappa^{\mbox{\tiny{eff}}}\equiv\kappa_{0}+\kappa_{1},
\label{fp-les-chiusa}
\end{equation}
where $\kappa_1$ is a constant depending on the flow properties that
can be explicitly calculated within the Kraichnan model. 
Eq.~(\ref{fp-les-chiusa}) is just the long-known 
{\it constant eddy-diffusivity closure}.

At the next-leading order we find 
\begin{equation}
\partial_t\tilde{\theta}+\tilde{\boldsymbol{v}}\cdot\nabla\tilde{\theta}=
\kappa_{\alpha\beta}^{T}\nabla_{\alpha}\nabla_{\beta}\tilde{\theta}
+\tilde{f}\;,
\label{effettiva}
\end{equation}
with $\kappa_{\alpha\beta}^{T}(\boldsymbol{x},t)=
\delta_{\alpha\beta}\kappa^{\mbox{\tiny{eff}}}- a l^2 \,e_{\alpha\beta}$,
and
$e_{\alpha\beta}=1/2\,(\nabla_{\alpha}\tilde{v}_{\beta}+\nabla_{\beta}\tilde{v}_{\alpha})$.
The filter dependent factor $a$ can be determined analytically
within the Kraichnan model. 
Equation (\ref{effettiva}) is the passive scalar 
analog (see, \eg, \cite[Kang \& Meneveau  2001]{KM01})
of the so-called ``mixed-model''
(nonlinear closure plus scalar eddy-viscosity)
used in Navier-Stokes turbulence (see, \eg, 
\cite[Borue \& Orszag  1998]{BO98}). Such mixed model was also
invoked by  \cite[Kang \& Meneveau (2001)]{KM01} to reproduce the 
correct amount of anisotropy in heated turbulent jets. \\
The nonlinear closure can be derived also starting from
a  Taylor expansion in the spirit of \cite[Leonard (1974)]{L74}
performed on a modified Leonard term (see, \eg, \cite[Horiuti 1997]{H97}).
A purely dissipative effective viscosity model 
is usually added because the sole
nonlinear model does not suffice, its dissipation being far too low.
Here, the constant eddy-diffusivity plus the nonlinear model follow
from first-principles without additional requirements.
The turbulent eddy-diffusivity 
$\kappa^{T}$ depends on the coarse-grained velocity field: 
for this reason we will dub this model
{\it dynamical eddy-diffusivity closure}.

\begin{figure}
\begin{center}
\includegraphics{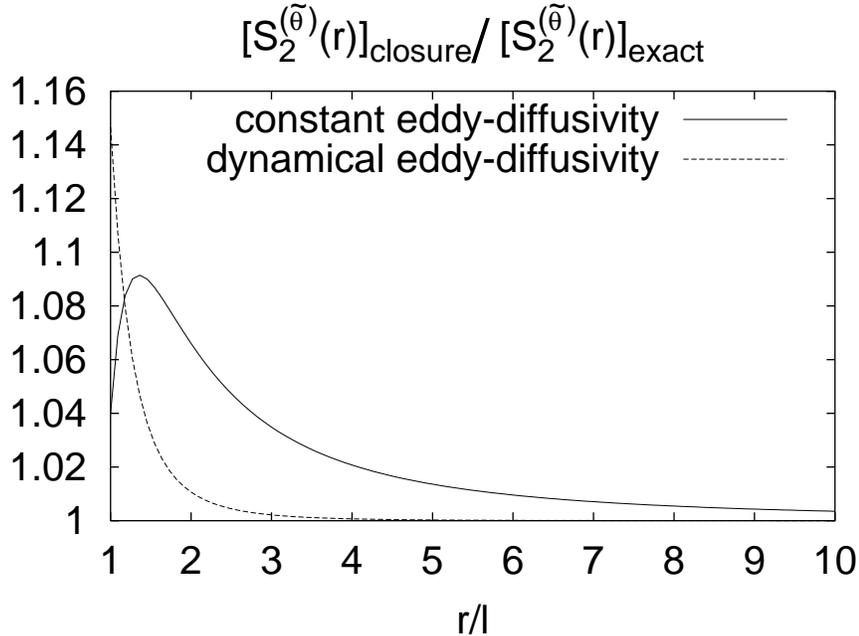}
\caption{The coarse-grained structure
functions obtained from the constant eddy-diffusivity closure, 
and the dynamical eddy-diffusivity model, normalized with respect 
to the exact filtered structure function (\ref{s2-large}).}
\label{fig:test}
\end{center}
\end{figure}

In order to compare the performances of the various models 
we will explicitly compute the structure function 
$S_2^{(\tilde{\theta})}(r)= 
\langle [\tilde{\theta}(\boldsymbol{x}+\boldsymbol{r},t)
-\tilde{\theta}(\boldsymbol{x},t)]^2 \rangle$ of the  coarse-grained
scalar field according to Eq.~(\ref{fp-les-chiusa}) 
or to Eq.~(\ref{effettiva}) and compare them with the exact value 
obtained from the solution of Eq.~(\ref{fp}) upon filtering.
The result is shown in Fig.~\ref{fig:test}: the dynamical 
eddy-diffusivity closure gives a structure function that is almost 
indistinguishable from the exact result already at scales $r \approx 2\,l$ 
whereas the constant eddy-diffusivity is not very effective in the range 
$r \lesssim 10\, l$. 

\section{A systematic approach to LES closure}
\label{systematic}
\subsection{The Kraichnan model}
In order to carry out our analysis, we need to specialize
Eq.~(\ref{fp}) to a class of random velocity
fields. To be more specific, the velocity field is
assumed here to be Gaussian,  of zero mean, statistically
stationary, homogeneous and isotropic, $\delta$-correlated in time and
with inertial-range power-law behavior. Its statistics is fully determined by
 the correlation function
\begin{equation}
\label{correlation}
\langle
\left [v_{\alpha}(\boldsymbol{x}_1,t)-v_{\alpha}(\boldsymbol{x}_2,t)\right ]
\left [v_{\beta}(\boldsymbol{x}_1,0)-v_{\beta}(\boldsymbol{x}_2,0)\right
]\rangle= 
2 D_{\alpha\beta}^{(\boldsymbol{v})}(\boldsymbol{x}_1 -\boldsymbol{x}_2)
\delta(t),
\end{equation}
where $D_{\alpha\beta}^{(\boldsymbol{v})}(\boldsymbol{r})\equiv
D_0\,r^{\xi}[(d+\xi-1)\delta_{\alpha\beta}-\xi r_{\alpha}
r_{\beta}/r^2]$,
$r\equiv |\boldsymbol{r}|=|\boldsymbol{x}_2-\boldsymbol{x}_1|$ and
$d$ is the space dimension. The assumption of
$\delta$-correlation in time is of course far from the reality, but it
has the remarkable feature of leading to closed equations for
equal-time correlation functions
$C^{(\theta)}_{n}\equiv\left\langle\theta(\boldsymbol{x}_1,t)\ldots\theta
(\boldsymbol{x}_n,t)\right\rangle$ of any order $n$ (see, \eg,
\cite[Falkovich \etal\ 2001]{FGV01}).  The parameter $\xi$ governs
the roughness of the velocity field, whose H\"older exponent is $\xi/2$.
Due to the white-in-time character of the flow, the Kolmogorov
value is $\xi=4/3$. \\
A convenient choice 
for the forcing term is to take $f$ random, 
Gaussian, statistically homogeneous and
isotropic, white in time, of zero mean and with correlation function
\begin{equation}
\langle f(\boldsymbol{x}_1,t)\,f(\boldsymbol{x}_2,0)\rangle =
F\left(r/L\right)\,\delta(t)
\label{fcorr}
\end{equation}
with $F\left(r/L\right)$ decreasing rapidly for $r\gg L$.
Since $l \ll L$ we have $\tilde{f}\simeq f$.

In this framework the
expression for the second order correlation function $C^{(\theta )}_{2}$ 
can be derived analytically (\cite[Falkovich \etal\ 2001]{FGV01}).
In the inertial range $\eta \ll r \ll L$ 
the exact second-order structure function of the scalar field 
$S_2^{(\theta)}(r)\equiv\langle[\theta(\boldsymbol{x}_1,t)-
\theta (\boldsymbol{x}_2,t)]^2\rangle$ reads:
\begin{equation}
S^{(\theta )}_2(r)=
\frac{2 F(0)}{\zeta_2(d-1)d D_0}\, r^{\zeta_2}\qquad \zeta_2=2-\xi,
\label{s2}
\end{equation}
where $F(0)$, defined in (\ref{fcorr}), is the average injection
rate of scalar variance.  
The exponent $\zeta_2$ coincides with the predictions based on
dimensional arguments, and is $\zeta_2=2/3$ for 
$\xi=4/3$, according to the Kolmogorov--Obukhov--Corrsin scaling.

From now on, we will confine ourselves to space dimension $d=3$ and specialize
$G_l$ to a top-hat filter, that is $G_l(r)=3/(4\pi l^3)$ if 
$r<l$ and $0$ otherwise.

To provide a benchmark for the various closures, we first evaluate
the exact value of the 
coarse-grained structure function $S^{(\tilde{\theta})}_2\equiv
\langle[\tilde{\theta}(\boldsymbol{x}_1,t)-\tilde{\theta}
 (\boldsymbol{x}_2,t)]^2\rangle$. 
A double integration on (\ref{s2}) yields
\begin{equation}
[S^{(\tilde{\theta})}_2(r)]_{\textrm{\tiny exact}}= 
S^{(\theta )}_2(r)\left [1+\frac{(2-\xi)(3-\xi)}{5}\left
(\frac{l}{r}\right )^2+O\left (\frac{l}{r}\right )^4\right ]\;.
\label{s2-large}
\end{equation}
Clearly, as the separation $r$ increases and gets far apart 
from the filter scale $l$
the unfiltered result is recovered. 
Equation (\ref{s2-large}) represents therefore the best result that can be 
 achieved  by means of a closure. 

\subsection{Exact statistical equations for unfiltered and filtered fields} 

As a first step, we derive the exact equations for the two-point correlation
function of the filtered and of the unfiltered field.
It is more convenient to start the analysis 
from Eq.~(\ref{fp}) by substituting
$\boldsymbol{v}=\tilde{\boldsymbol{v}}+\boldsymbol{v}'$ and
$\theta=\tilde{\theta}+\theta '$ in the advective term 
$\boldsymbol{v}\cdot\nabla\,\theta$.  
Equation (\ref{fp}) takes then the form:
\begin{equation}
\partial_t \theta (\boldsymbol{x}_2,t)
+\tilde{\boldsymbol{v}}(\boldsymbol{x}_2,t)
\cdot\nabla\,
\tilde{\theta}(\boldsymbol{x}_2,t)
=\kappa_0\Delta\theta (\boldsymbol{x}_2,t) +
f (\boldsymbol{x}_2,t) -{\cal S}(\boldsymbol{x}_2,t)  \;,
\label{fp-work1}
\end{equation}
from which we can immediately derive the equation for
$C^{(\theta )}_2(r,t)=\langle \theta(\boldsymbol{x}_1,t) 
\theta(\boldsymbol{x}_2,t) \rangle$:
\begin{equation}
\partial_t C^{(\theta )}_2(r,t) + 2\langle\theta
(\boldsymbol{x}_1,t) 
\tilde{\boldsymbol{v}}\cdot\nabla\tilde{\theta}(\boldsymbol{x}_2,t)
\rangle - 2\kappa_0\Delta C^{(\theta )}_2(r,t) 
=F(r)-2\langle \theta(\boldsymbol{x}_1){\cal S}(\boldsymbol{x}_2)\rangle\;.
\label{work1}
\end{equation}
A double convolution of the above equation with the filter $G_l$
yields the exact equation for the correlation of the filtered field:
\begin{equation}
\partial_t C^{(\tilde{\theta})}_2(r,t) + 2\langle\tilde{\theta}
(\boldsymbol{x}_1,t) 
\widetilde{\tilde{\boldsymbol{v}}\cdot\nabla\tilde{\theta}}(\boldsymbol{x}_2,t)
\rangle - 2\kappa_0\Delta C^{(\tilde{\theta})}_2(r,t)
=F(r)-2\langle \tilde{\theta}(\boldsymbol{x}_1)
\tilde{\cal S}(\boldsymbol{x}_2)\rangle\;.
\label{c2-les1}
\end{equation}
This is the starting point for our systematic procedure to build 
closure approximations. It contains two terms, the second one in the 
left hand side (l.h.s.)
and the last one in the right hand side (r.h.s.), 
which are not expressed as functions
of large-scale fields only. In the following, we will find approximate
closed expressions for the unclosed terms perturbatively in $l/r$. 

\subsection{Constant eddy-diffusivity closure}

In order to calculate 
the second term in the l.h.s. of  Eq.~(\ref{c2-les1}),
let us start from 
$\langle \theta (\boldsymbol{x}_1)
{\cal S}(\boldsymbol{x}_2)\rangle$. Its expression can be easily obtained 
exploiting, \eg,
the Furutsu--Novikov's functional Gaussian integration
(see, \eg, \cite[Frisch 1995]{F95}), which holds for Gaussian velocities and
forcings. It reads:
\begin{equation}
\label{nuova}
\langle \theta(\boldsymbol{x}_1){\cal S}(\boldsymbol{x}_2)\rangle=
\frac{F(0)2^{\xi+2} (3-\xi)}{(4+\xi)(6+\xi)} 
\left (\frac{l}{r}\right )^{\xi}
-\frac{F(0)}{30}\xi^2 \left (\frac{l}{r}\right
)^2+O\left (\frac{l}{r}\right )^{2+\xi}\!\!\!\!.
\end{equation}
Our first claim is that, at order $(l/r)^{2}$, with an error
of order $(l/r)^{2+\xi}$ we have:
\begin{equation}
\label{ts}
\langle \theta(\boldsymbol{x}_1){\cal S}(\boldsymbol{x}_2)\rangle=
\langle \tilde{\theta}(\boldsymbol{x}_1)\tilde{\cal S}(\boldsymbol{x}_2)\rangle
=\langle \tilde{\theta}(\boldsymbol{x}_1) {\cal S}(\boldsymbol{x}_2)\rangle 
\end{equation}
and
\begin{equation}
\langle\tilde{\theta}
(\boldsymbol{x}_1,t) 
\widetilde{\tilde{\boldsymbol{v}}\cdot\nabla\tilde{\theta}}(\boldsymbol{x}_2,t)
\rangle=\langle\tilde{\theta}
(\boldsymbol{x}_1,t) 
\tilde{\boldsymbol{v}}\cdot\nabla\tilde{\theta}(\boldsymbol{x}_2,t)\rangle.
\label{leonard0}
\end{equation}
The derivation of  (\ref{ts}) and  (\ref{leonard0}) is postponed to
the Appendix \ref{appendix}.
Before proceeding further, some comments on (\ref{leonard0}) are 
in order.
In plain words, (\ref{leonard0}) tells us that the 
 Leonard-type term does not contribute, at order $O((l/r)^2)$, 
to the equation for the 
second-order coarse-grained scalar correlation function. Since our closures
are derived from the latter equation, it follows that 
the Leonard-type term will not contribute to
small-scale parameterizations. This fact is not a consequence of 
the Kraichnan advection model but rather seems to hold for general advection
models. Indeed, a standard 
expansion in the spirit of \cite[Leonard (1974)]{L74}
(see also, \eg, \cite[Horiuti 1997]{H97}) performed on
$\langle\tilde{\theta}(\boldsymbol{x}_1) {\cal L}(\boldsymbol{x}_2) 
\rangle$ with ${\cal L}$ given by 
(\ref{leo-stress}), yields at the lowest-order in the filter width 
the expression
\begin{equation}
\langle\tilde{\theta}(\boldsymbol{x}_1) {\cal L}(\boldsymbol{x}_2) 
\rangle \sim l^2 \Delta \langle\tilde{\theta}
(\boldsymbol{x}_1,t) 
\tilde{\boldsymbol{v}}\cdot\nabla\tilde{\theta}(\boldsymbol{x}_2,t)\rangle
.
\end{equation}
The latter expression is trivially zero, since
$\langle\tilde{\theta}
(\boldsymbol{x}_1,t) 
\tilde{\boldsymbol{v}}\cdot\nabla\tilde{\theta}(\boldsymbol{x}_2,t)\rangle$
is the flux of scalar variance which is independent of 
$r=|\boldsymbol{x}_2-\boldsymbol{x}_1|$ 
provided that $r$ falls in the inertial range of scales. \\
For standard closure models based on single-point quantities, 
the contribution from the Leonard stress in the parameterizations
is, generally speaking, nonzero.

Let us now focus on consequences of relations (\ref{ts}) and (\ref{leonard0})
on Eq.~(\ref{c2-les1})  evaluated at order $(l/r)^{\xi}$.
At this order, from (\ref{nuova}) and (\ref{ts}) we have:
\begin{equation}
-2\langle \tilde{\theta}(\boldsymbol{x}_1)\tilde{\cal S}(\boldsymbol{x}_2)\rangle=
2\kappa_1 \Delta C^{(\tilde{\theta})}_2(r,t).
\label{leonard}
\end{equation}
The above expression follows by comparing
(\ref{nuova}) at the order $(l/r)^{\xi}$
with the contribution coming from the 
diffusive term, $2\kappa_0\Delta C^{(\tilde{\theta})}_2(r,t) =
-F(0)\kappa_0(3-\xi)/(3 D_0 r^{\xi})$.
We immediately 
realize that the term of order $(l/r)^{\xi}$ in 
(\ref{nuova}) corresponds to an effective
diffusive term with an eddy diffusivity $\kappa_1\equiv l^{\xi}
[2^{\xi} 24 D_0]/[(\xi+4)(\xi+6)]$. This gives rise to an effective
dissipative scale comparable to $l$: 
\begin{equation}
\eta^{\mbox{\tiny{eff}}}\equiv \left [(2 l)^{\xi}\left ( \frac{24}{(\xi+4)(\xi+6)}
\right) + \eta^{\xi}_0 \right ]^{1/\xi},
\end{equation}
where
$\eta_0\equiv(\kappa_0/D_0)^{1/\xi}$ is the molecular dissipation scale.\\
At order $(l/r)^{\xi}$, Eq.~(\ref{c2-les1}) is thus closed
in the large scales fields and, moreover, due to (\ref{leonard0})
it has the same structure of
the equation for $C^{(\theta )}_2$ but with an effective diffusivity, 
$\kappa_0+\kappa_1$. 
Eq.~(\ref{fp-les}) thus takes the form:
\begin{equation}
\partial_t \tilde{\theta}+\tilde{\boldsymbol{v}}\cdot\nabla\,
\tilde{\theta}=\kappa^{\mbox{\tiny{eff}}}\Delta\tilde{\theta} +
\tilde{f}\qquad \kappa^{\mbox{\tiny{eff}}}\equiv\kappa_{0}+\kappa_{1}.
\label{cedc}
\end{equation}

Starting from (\ref{fp-les-chiusa}) one can  deduce the equation
for the correlation function $C_2^{(\tilde{\theta})}$ exploiting
the Furutsu--Novikov's functional Gaussian integration. 
When doing that, the next step will consist in comparing
the resulting expression for
$S_2^{(\tilde{\theta})}$ 
 with (\ref{s2-large}).
After some simple but quite lengthy algebra, the calculation leads to:
\begin{equation}
\!\!\!\!S^{(\tilde{\theta})}_2= S^{(\theta )}_2\left [1+\frac{(2-\xi)(3+\xi)}{5}\left
(\frac{l}{r}\right )^2+ O\left (\frac{l}{r}\right )^4\right ].
\label{s2-eddy}
\end{equation}
By comparison with (\ref{s2-large}), the above expression
permits to quantify the error, which occurs at order $(l/r)^2$,
on the second order structure
function due to the closure (\ref{cedc}). 
The degree of accuracy of the constant eddy-diffusivity 
description can be perceived by looking at  Fig.~\ref{fig:test},
obtained for $\xi=4/3$, corresponding to the Kolmogorov scaling 
law for the advecting velocity field.

\subsection{Dynamical eddy-diffusivity closure}

Our aim is now to improve the eddy-diffusivity closure, exact 
at order $(l/r)^{\xi}$, by introducing a new closure which is accurate
up to order $(l/r)^2$. 
It is not difficult, although quite lengthy, to verify that 
the large-scale equation has the form (\ref{effettiva}):
\begin{equation}
\partial_t\tilde{\theta}+\tilde{\boldsymbol{v}}\cdot\nabla\tilde{\theta}=
\kappa_{\alpha\beta}^{T}\nabla_{\alpha}\nabla_{\beta}\tilde{\theta}
+\tilde{f}\qquad \kappa_{\alpha\beta}^{T}(\boldsymbol{x},t)=
\delta_{\alpha\beta}\kappa^{\mbox{\tiny{eff}}}- a l^2 \,e_{\alpha\beta}\;,
\label{dedc}
\end{equation} 
with
$e_{\alpha\beta}=1/2\,(\nabla_{\alpha}\tilde{v}_{\beta}+\nabla_{\beta}\tilde{v}_{\alpha})$.\\
The filter-dependent factor $a= \int d^3 \boldsymbol{r}
G_l(r) r^2/(3 l^2)$.
For the top-hat filter one immediately obtains $a=1/5$. \\

The equation for $\langle \tilde\theta^2\rangle$, deduced from (\ref{dedc}),
at the statistically stationary state, 
reads (remember that $\tilde{f} = f$):
\begin{equation}
F(0)=2\kappa^{\mbox{\tiny{eff}}}\langle (\nabla\tilde\theta)^2\rangle
-a l^2 \langle e_{\alpha\beta} \nabla_{\alpha}\tilde{\theta}
\nabla_{\beta}\tilde{\theta}   \rangle
\label{bala1}
\end{equation}
which states the energy balance between production
(controlled by $F(0)$) and dissipation.\\
The fact that the first term on the r.h.s. of (\ref{bala1})
gives a  dissipative contribution is evident. This is actually the
case also for the second term. To show that, let us start from 
simple physical considerations.  Transforming to the principal
coordinates, $\boldsymbol{x}'$, of $e_{\alpha\beta}$, the term
$-a l^2  e_{\alpha\beta} \nabla_{\alpha}\tilde{\theta}
\nabla_{\beta}\tilde{\theta}  $ becomes:
\begin{equation}
-a l^2 (c_1 \nabla_{1'}\tilde{\theta}\nabla_{1'}\tilde{\theta}+ 
c_2 \nabla_{2'}\tilde{\theta}\nabla_{2'}\tilde{\theta} + 
c_3 \nabla_{3'}\tilde{\theta}\nabla_{3'}\tilde{\theta}).
\label{euristic}
\end{equation}
Because of incompressibility one has $c_1+c_2+c_3=0$, and
stretching in a given direction is always accompanied by compression
along, at least, another direction. By virtue of the fact that
strong scalar gradients are expected to be aligned along the
direction of maximum compression (corresponding to negative $c_i$),
it then follows that, on average, (\ref{euristic}) is expected 
to be positive. This is the mechanism which leads to the well-known
ramp-and-cliff structure observed in passive scalar turbulence
both for Navier-Stokes velocity fields and for the Kraichnan ensemble
(see, e.g., \cite[Celani \etal\ 2001]{CLMV01}).\\
The above arguments can be substantiated  within the 
Kraichnan model. Indeed, exploiting the Furutsu--Novikov's functional 
Gaussian integration by parts, one obtains:
\begin{equation}
-a l^2\langle  e_{\alpha\beta} \nabla_{\alpha}\tilde{\theta}
\nabla_{\beta}\tilde{\theta} \rangle = a \frac{l^2}{6} 
\langle(\nabla_i \tilde{v}_j)^2\rangle \langle (\nabla\tilde\theta)^2\rangle.
\label{vero}
\end{equation}
However, it is clearly possible to observe, locally in space and time, 
positive values of $e_{\alpha\beta} \nabla_{\alpha}\tilde{\theta}
\nabla_{\beta}\tilde{\theta}  $, or, in other words, backscattering events
responsible for negative contributions to the scalar energy flux
(see, \eg, \cite[Borue \& Orszag  1998]{BO98} for discussions 
on backscattering events in hydrodynamics turbulence).

Focusing now on the unfiltered field, the  
balance equation has the well-known form:
\begin{equation}
F(0)=2\kappa_0\langle (\nabla\theta)^2\rangle .
\label{bala2}
\end{equation}
Equating the l.h.s.'s of (\ref{bala1}) and (\ref{bala2}),
and recalling the inequality 
$\kappa^{\mbox{\tiny{eff}}}+l^2/60 \langle(\nabla_i
\tilde{v}_j)^2\rangle \\ \gg \kappa_0$,
one concludes that the gradients of the large scale
scalar field are smaller than the gradients of the unfiltered
field. This is consistent with an effective dissipative scale
comparable to $l$ and thus much larger than $\eta$ by definition.

To prove Eq.~(\ref{dedc}), let us start by rewriting 
Eq.~(\ref{c2-les1}) with
$\langle \tilde{\theta}(\boldsymbol{x}_1)
\tilde{\cal S}(\boldsymbol{x}_2)\rangle$  
expressed in terms of (\ref{nuova}). Using (\ref{ts}) we obtain:
 \begin{equation} 
\label{eq:1}
\partial_t C_2^{(\tilde{\theta})}(r,t)+2\langle
\tilde{\theta}(\boldsymbol{x}_1,t)\tilde{\boldsymbol{v}}\cdot\nabla
\tilde{\theta}(\boldsymbol{x}_2,t)\rangle  
-2\kappa^{\mbox{\tiny{eff}}}\Delta C_2^{(\tilde{\theta})}(r,t)-F(r)=
\frac{F(0)\xi^2}{15}\left(\frac{l}{r}\right)^2
\end{equation}
where the contribution of order $(l/r)^{\xi}$ in (\ref{nuova})
has been incorporated  in the 
eddy-diffusivity term.
 From (\ref{effettiva}) we immediately obtain the equation for 
$C_2^{(\tilde{\theta})}$:
\begin{eqnarray} 
\partial_t C_2^{(\tilde{\theta})}(r,t)&+&2\langle
\tilde{\theta}(\boldsymbol{x}_1,t)\tilde{\boldsymbol{v}}\cdot\nabla
\tilde{\theta}(\boldsymbol{x}_2,t)\rangle
- 2\kappa^{\mbox{\tiny{eff}}}\Delta C_2^{(\tilde{\theta})}(r,t)-F(r)
=\nonumber\\
&-& 2 a l^2 \langle\tilde{\theta}(\boldsymbol{x}_1)
e_{\alpha\beta}\nabla_{\alpha}\nabla_{\beta}\tilde{\theta}(\boldsymbol{x}_2) 
\rangle + O[(l/r)^{2+\xi}] \;,
\label{eq:1bis}
\end{eqnarray}
with $a=1/5$.
We finally have to show that the r.h.s. of (\ref{eq:1})
and the r.h.s. of (\ref{eq:1bis}) coincide up to order $(l/r)^2$.
In order to evaluate the r.h.s. of (\ref{eq:1bis}) we have
to exploit again the Furutsu--Novikov's functional Gaussian integration
by parts. One thus needs to compute the functional derivative 
$\delta \tilde{\theta}(\boldsymbol{x},t) / \delta
\tilde{\boldsymbol{v}}(\boldsymbol{x}'',t'')$ which can be easily obtained 
from (\ref{effettiva}). Accounting for the delta-correlation in time and 
utilizing the expansions
\begin{equation}
\label{v2es2tilde}
D_{\alpha\beta}^{(\tilde{\boldsymbol{v}})}(\boldsymbol{r})=
D_{\alpha\beta}^{(\boldsymbol{v})}(\boldsymbol{r})
\left\{1+O\left[\left(\frac{l}{r}\right)^{\xi}\right]\right\} \qquad
 S_2^{(\tilde{\theta})}(r)=S_2^{(\theta)}(r)\left\{
1+O\left[\left(\frac{l}{r}\right)^2\right]\right\},
\end{equation}
one ends up exactly with the r.h.s. of (\ref{eq:1}).
Exploiting once more the Furutsu--Novikov's functional Gaussian integration
by parts, it is not difficult (although quite lengthy) to verify that
the expression  (\ref{s2-large}) for $S^{(\tilde{\theta})}_2$ is
obtained from (\ref{effettiva}). The remaining error is pushed at order 
$(l/r)^4$.

\section{Conclusions and perspectives}
\label{conclusions}

Summarizing, a systematic procedure to derive closed dynamical equations
for a coarse-grained passive scalar field in the statistical steady-state
has been obtained in the framework of the Kraichnan advection model.

The question that naturally arises, is whether those results are relevant
to realistic  advection models. The answer is given by the
outcome of the procedure itself. We recover from first principles 
two well-known closures that are commonly used in applications:
the constant eddy-diffusivity parameterization of small-scales,
and the passive scalar version of the nonlinear eddy-viscosity closure
used in hydrodynamic turbulence. Of course, the value of the effective
diffusivity $\kappa^{\mbox{\tiny eff}}$ and of the numerical parameter
$a$ that appear in these closures can be analytically computed
only in the Kraichnan model. 
However, we believe that the form of the
parameterization can be exported without modifications to real
situations as well. Clearly, in this case the free parameters
(\eg $\kappa^{\mbox{\tiny eff}}$ and $a$) have to be determined
{\it a posteriori} by some empirical procedure. The validity of
this approach can be checked by direct numerical simulations.

Let us conclude by mentioning a possible generalization of our work.
Our analysis has been carried for the second-order correlation function
of the scalar field. There are two reasons for this choice. First,
the second order correlation function is the Fourier transform of
the spectrum of scalar variance, a widely used statistical
indicator to characterize most of the statistical properties of
scalar turbulence. Second,
for the Kraichnan model only the second-order correlation function
has  a simple, closed analytical expression. 
For higher-order correlation functions
only perturbative expressions (for example in the limit of small
$\xi$)  are available (see \cite[Falkovich \etal\ 2001]{FGV01}).
However, should have we focused on  a higher-order correlation function,
how would our results change~?
Although  the analysis appears much more 
 cumbersome than the one presented here, the procedure described in
\S\,\ref{systematic} can be completed as well:
it is still possible to obtain a closed equation for the coarse-grained
 correlation function at any order in $l/r$, from which one can identify 
the corresponding dynamical equations for the large-scale scalar field.
The question is: will the latter dynamical equation have the same structure 
of the coarse-grained scalar equation derived from the
second-order correlation~? And if this is the case, will the coefficients
be the same~?
Even if the functional form of the closure is preserved, a
 modification of the effective coefficients would mean
that strong small-scale fluctuations  
-- associated to higher-order correlation functions -- 
must be described by parameters different from the ones used for less intense
fluctuations.
That would question the applicability of closure 
models to the description of the statistics of turbulent fields
as temperature or concentration, which are
characterized by a wide range of fluctuation intensities.
This challenging issue is left
for future research.

\begin{acknowledgments}
This work has been supported by Cofin 2001, prot. 2001023848
and by HPRN-CT-2002-00300 ``Fluid Mechanical Stirring and Mixing''. 
We acknowledge useful discussions with Alessandra Lanotte.
\end{acknowledgments}

\appendix
\section{Proof of Eq.~(\ref{ts}) and Eq.~ (\ref{leonard0})}\label{appendix}

Let us write, from (\ref{nuova}),
$\langle \theta(\boldsymbol{x}_1){\cal S}(\boldsymbol{x}_2)\rangle=
A (l/r)^{\xi} +B(l/r)^{2}$. By  direct calculation it is easily checked that,
for any filter $G_l$ (normalized and isotropic), the following 
relations hold up to the second order in $l/r$.
 \begin{eqnarray}
&&\langle\tilde\theta(\boldsymbol{x}_1){\cal S}(\boldsymbol{x}_2)\rangle=
\int d^3\boldsymbol{s}G_l(s)\langle\theta(\boldsymbol{x}_1+\boldsymbol{s})
{\cal S}(\boldsymbol{x}_2)\rangle= \nonumber \\
&& \int d^3\boldsymbol{s}G_l(s)\left[A\left(\frac{l}{|\boldsymbol{r}+\boldsymbol{s}|}
\right)^\xi+B\left(\frac{l}{|\boldsymbol{r}+\boldsymbol{s}|}\right)^2\right]+
O\left(\frac{l}{r}\right)^{\xi+2}= \nonumber\\
&&A\left(\frac{l}{r}\right)^\xi\int ds s^2 G_l(s)\int_{-1}^1
 d(\cos\vartheta)\int_0^{2\pi}d\varphi 
\left[1-\xi\cos\vartheta\frac{s}{r}+O\left(\frac{s}{r}\right)^2\right]+
\nonumber
\\
&&B\left(\frac{l}{r}\right)^2\int ds s^2 G_l(s)\int_{-1}^1 
d(\cos\vartheta) \int_0^{2\pi}d\varphi
\left[1-2\cos\vartheta\frac{s}{r}+O\left(\frac{s}{r}\right)^2\right]+
O\left(\frac{l}{r}\right)^{\xi+2}=
\nonumber\\
&&A\left(\frac{l}{r}\right)^\xi+B\left(\frac{l}{r}\right)^2+O\left(\frac{l}{r}\right)^{\xi+2}=
\langle\theta(\boldsymbol{x}_1){\cal S}(\boldsymbol{x}_2)\rangle+
O\left(\frac{l}{r}\right)^{\xi+2}\;.\nonumber
 \end{eqnarray}
Similarly, convolving also over $\boldsymbol{x}_2$, one finds:
\begin{equation}
\langle\tilde\theta(\boldsymbol{x}_1)\tilde{\cal S}(\boldsymbol{x}_2)\rangle=
\langle\tilde\theta(\boldsymbol{x}_1){\cal S}(\boldsymbol{x}_2)\rangle
+O\left(\frac{l}{r}\right)^{\xi+2}
=\langle\theta(\boldsymbol{x}_1){\cal S}
(\boldsymbol{x}_2)\rangle+O\left(\frac{l}{r}\right)^{\xi+2}\;.
\end{equation}

To prove (\ref{leonard0}),
let us now consider (\ref{c2-les1})
and the equation obtained from (\ref{work1})
with the sole convolution over $\boldsymbol{x}_1$.
 Because of stationarity all time derivatives  vanish
 and for $r$ in the inertial range 
the two-point terms proportional to molecular 
diffusivity $\kappa_0$ are negligible.
We can thus rewrite the above two equations as:
\begin{equation}
2\langle\tilde{\theta}(\boldsymbol{x}_1)\widetilde{\tilde{\boldsymbol{v}}\cdot\nabla\tilde{\theta}}(\boldsymbol{x}_2)\rangle=F(r)
-2\langle\tilde{\theta}(\boldsymbol{x}_1)\tilde {\cal S}(\boldsymbol{x}_2)\rangle
\label{passo1}
\end{equation}
\begin{equation}
2\langle\tilde{\theta}(\boldsymbol{x}_1)\tilde{\boldsymbol{v}}\cdot\nabla\tilde{\theta}(\boldsymbol{x}_2)\rangle=F(r)-2\langle\tilde{\theta}(\boldsymbol{x}_1){\cal S}(\boldsymbol{x}_2)\rangle\;.
\label{passo2}
\end{equation}
 Subtracting (\ref{passo2}) from (\ref{passo1}), and using (\ref{ts}), we
 conclude that
 $\langle\tilde{\theta}{\cal L}\rangle=O[(l/r)^{\xi+2}]$. This proves
 (\ref{leonard0}).

\end{document}